\documentclass{article} 

\usepackage{amsmath,amssymb,amsthm,latexsym} 

\usepackage{stmaryrd,wasysym,upgreek,mathrsfs,dsfont} 
\usepackage[english]{babel} 
\usepackage{graphicx,color} 

\usepackage{array}

\newtheorem{lemma}{Lemma}[section]

\newtheorem{theorem}{Theorem}[section]
\newcommand{\be}{\begin{equation}}
\newcommand{\ee}{\end{equation}}
\newcommand{\bea}{\begin{eqnarray}}
\newcommand{\eea}{\end{eqnarray}}

\newcommand{\Tr}{{\rm Tr}}

\newcommand{\prf}{{\noindent {\rm \bf Proof}\, }}

\begin{document} 

 \title{Note on the Intermediate Field Representation\\
of $\phi^{2k}$ Theory in Zero Dimension}
\author{Luca Lionni and Vincent Rivasseau\\
Laboratoire de Physique Th\'eorique, CNRS UMR 8627,\\ 
Universit\'e Paris XI,  F-91405 Orsay Cedex, France}

\maketitle 
\begin{abstract} 
This note is a sequel to \cite{Rivasseau:2010ke}.
We correct  the intermediate field representation
for the stable $\phi^{2k}$ field theory in zero dimension introduced there and extend it to the case of complex conjugate fields.
For $k=3$ in the complex case we also provide an improved representation which relies on 
ordinary convergent Gaussian integrals rather than oscillatory integrals.
\end{abstract} 

\begin{flushright}
LPT-20XX-xx
\end{flushright}
\medskip


\medskip

\section{Introduction}

The intermediate field (hereafter called IF) representation and the associated constructive loop vertex expansion (LVE)
\cite{R1}-\cite{Rivasseau:2013ova} 
have been increasingly used in recent years 
\cite{MNRS}-\cite{Lahoche:2015zya} for models with quartic interactions. 
It is important to extend such techniques to models with higher order stable interactions,
as first attempted in \cite{Rivasseau:2010ke}. The case of a $\phi^6$ interaction is treated in sections 2 and 3 of
\cite{Rivasseau:2010ke}, using imaginary Gaussian measures with a small contour deformation, and the case
of a general $\phi^{2k}$ interaction is sketched in section 4 of \cite{Rivasseau:2010ke}. 
Unfortunately Lemma 4.1 as stated there is not correct and requires a slight modification. Also the number of intermediate fields introduced in \cite{Rivasseau:2010ke} is not optimal. Finally we found that the loop vertex expansion in \cite{Rivasseau:2010ke}
is not correct since the interpolation of imaginary Gaussian covariances through forest formulas is 
not fully justified. Hence the main purpose of this paper is to correct \cite{Rivasseau:2010ke}, to give a proof of Borel-Leroy summability 
(since the one in \cite{Rivasseau:2010ke} is incorrect), 
and to give improved intermediate field representations 
for the partition functions of such toy models. It should lay the ground for 
future extensions to high order interactions of constructive techniques such as the LVE or its multiscale extension \cite{Gurau:2013oqa}.

In the next section we gather some mathematical prerequisites on Gaussian imaginary integrals and Borel-Leroy summability. We
also introduce the models discussed in this paper, namely the (stable) $\lambda \phi^{2k}$ model
and its complex $\lambda (\bar \phi \phi)^{k}$ version. They are zero-dimensional, hence toy models 
useful to test constructive methods in quantum field theory \cite{Rivasseau:2009pi}.
Their partition functions $Z_k (\lambda)$ and $Z^c_k (\lambda)$ are the generating
functions for counting $\phi^{2k}$ or  $(\bar \phi \phi)^{k}$ vacuum Feynman graphs.  
Using their ordinary integral representation 
we check that  these partition functions are the Borel-Leroy sum of their perturbative expansion in powers of $\lambda$.

In Section \ref{heuristic}, starting
from this ordinary representation, we guess the form of an intermediate field representation,
using as in \cite{Rivasseau:2010ke} imaginary Gaussian integrals with  suitable
integration contours ensuring convergence. 
This guess is based on commuting some integrals without caring about convergence.

Then in Section \ref{analyt} we check that our guess is in fact
an absolutely convergent integral, which is again the Borel sum of its perturbative expansion.
Since this expansion is the same as the initial one,
from unicity of the Borel-Leroy sum we conclude \emph{a posteriori} that our guess
is indeed a correct (non-perturbative) representation of the partition functions $Z_k (\lambda)$ and $Z^c_k (\lambda)$.

The ``free energies" $\log Z_k (\lambda)$ and $\log Z^c_k (\lambda)$ are physically more interesting than the partition functions. They are the generating functions of \emph{connected}  $\phi^{2k}$ or  $(\bar \phi \phi)^{k}$ vacuum Feynman graphs.
The LVE combines an intermediate field representation with a forest formula and a replica trick to
compute directly these functions through a convergent expansion.
In spite of several attempts, we have not been able yet to define a convergent LVE for 
the imaginary Gaussian intermediate field representation of Section \ref{heuristic}, hence correct the last problem in \cite{Rivasseau:2010ke}. 
Therefore in Section 
\ref{improved} of this paper
we introduce still another intermediate representation, better adapted to this task. Since it is also 
more complicated, we limit ourselves to give it
in the $k=3$ complex case, leaving its generalization and detailed study to future works.

\section{Prerequisites}
\label{prereq}
\subsection{Imaginary Gaussian Measures}\label{imag}

Consider a function $f(z)$ which is analytic in the strip $\Im z \le \delta$ and exponentially bounded in that 
domain by $K  e^{\eta \vert z \vert} $ for some $0 \le  \eta  < \delta$, where $K$ is some constant.

\begin{figure}[!htb]
\centering
\includegraphics[width=10cm]{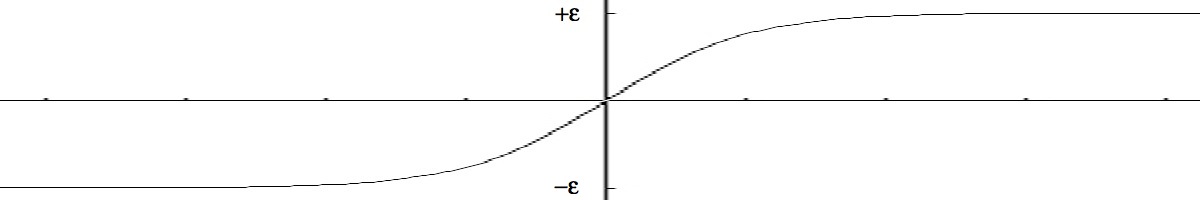}
\caption{The integration contour $C_{+,\epsilon}$.}
\label{contour+}
\end{figure}

The imaginary Gaussian integral of $f$ with covariance $\pm i  C$, where $C >0$, is defined as
\be  \int d \mu_{\pm i  C} (x) f (x)  := \int_{C_{\pm ,\epsilon}}  \frac{e^{- z^2  /\pm 2 i C} dz}{\sqrt{\pm 2 \pi i C}} f(z) = \int_{C_{\pm ,\epsilon}}  
\frac{e^{\pm iz^2  /2C} dz}{\sqrt{\pm 2 \pi i C}} f(z) \label{imgauss}
\ee
where the contour $C_{\pm , \epsilon}$
can be for instance chosen as $t \to z(t) =t \pm i \epsilon \tanh (t)$
for any $\epsilon \in ]C\eta, \delta [$, where $t  \in {\mathbb R}$.
Remark indeed that from our hypotheses on $f$, the integral \eqref{imgauss} is well defined 
and absolutely convergent for $C\eta < \epsilon < \delta$, and by Cauchy theorem, independent
of $\epsilon \in ]C\eta, \delta [$. The contour $C_{+, \epsilon}$ is shown in Figure \ref{contour+}.

Although the result of integration does not depend on the contour, actual bounds on the result typically depend on choosing 
particular contours in which $\epsilon$ is not too small, see Section \ref{analyt}. Furthermore the
Gaussian rules of integration still apply, e.g. defining $(2n -1)!! := (2n-1) (2n-3) \cdots 5.3.1$
\be  \label{regau} \int d \mu_{\pm i C}(x) x^{2n}=   (\pm i C)^n (2n-1)!! \ .
\ee
This is easy to check since a polynomial
is an entire function and we can deform the contour into $z = x+ ix$, in which case we recover an ordinary Gaussian integration.
Similarly
\be   \int d \mu_{\pm iC}(x) e^{ax}=   e^{\pm i Ca^2 /2}  , \label{contexp}
\ee
the integral being absolutely convergent for any contour such that $C\vert a \vert < \epsilon$.

The imaginary complex normalized Gaussian measures $ d \mu^c_{\pm i C}(z) $ 
of covariance $\pm i$ for a complex variable $z=x+iy$
is similarly defined as a pair of independent real normalized Gaussian measuresof covariance $C$, 
one for $x$ and one for $y$. \eqref{regau}-\eqref{contexp} generalize to 
\be   \int d \mu^c_{\pm iC}(x) (z \bar z)^{n}=   (\pm i C)^n  n! \label{imgau}
\ee
and
\be   \int d \mu^c_{\pm i C}(z) e^{a z + b \bar z }=   e^{\pm i a b C}  ,
\ee
again this last integral being absolutely convergent if $C\sup\{\vert a \vert, \vert b \vert \} < \epsilon$.
The integrals correspond to complexifications of the two dimensional integral $\int_{\mathbb C}dzd\bar z =\int_{-\infty}^{+\infty}dx\int_{-\infty}^{+\infty}dy$ into the product of two contour integrals on $C_{+, \epsilon}$, one for $x$
and one for $y$.

\subsection{Borel-Leroy summability}

Usual Borel summability concerns functions analytic in a domain with opening angle $\pi $ 
asymptotic to power series $\sum_{n=0}^\infty a_n\lambda^n$ with large $n$ behavior $a_n \simeq c^n n!$. 
Borel-Ler(oy summability extends to the case of functions with larger analyticity domains, of opening angle $k\pi $, 
but also worse asymptotic series with large $n$ behavior $a_n \simeq c^n[n!]^k$.  

Consider the Riemann surface ${\cal L}$ for the logarithm, namely the universal cover of ${\mathbb C}^\star$ with deck transformations
$(z,\theta) \to (z,\theta+2k\pi)$, $k \in {\mathbb Z}$. ${\cal L}$ can be embedded in ${\mathbb C}^\star \times {\mathbb R}$
since a point of ${\cal L}$ can be defined as a pair $(z,\theta)$ in ${\mathbb C}^\star \times {\mathbb R}$ 
such that $\theta$ is an argument of $z$. The logarithm is well-defined (single valued) on 
${\cal L}$. For $\rho >0$ we define the open domain
$D^k_\rho \subset {\cal L}$ by the equation $\Re \lambda^{- \frac{1}{k}}> \rho^{-1}$, 
which means $\Re [e^{  - \frac{1}{k} \log \lambda } ]  > \rho^{-1}$. $D^1_\rho$ is a disk
tangent to the imaginary axis of diameter $\rho$. For larger values of $k$
$D^k_\rho$ spreads over more and more sheets of ${\cal L}$ 
but remains close to the vertical axis (the origin $\lambda =0$). More precisely a point $\lambda$ is in $D^k_\rho$ iff its unique representative
$(z,\theta) \in {\mathbb C}^\star \times {\mathbb R}$ with  $\theta = \arg z\ (2 \pi)$
satisfies to $\vert \theta \vert < \frac{k \pi }{2}$  and $\vert z \vert^{1/k} <  \rho \cos \frac{\theta}{k}$ 
so it is at distance less than $\rho^k$ of the vertical axis $ \{0\} \times {\mathbb R}$ in $ {\mathbb C} \times {\mathbb R}$.

We note $R^N$ the $N$-th order Taylor remainder operator at the origin. It 
acts on a smooth function $f(\lambda)$ through
\bea  R^N  f = \lambda^{N}
\int_0^1 \frac{(1-t)^{N-1}}{(N-1)!} f^{(N)} ( t \lambda) dt .
\eea

\begin{theorem}

A power series $\sum_{n=0}^\infty a_n\lambda^n$ is Borel-Leroy summable of order $k$ 
to the function $f(\lambda)$  if the following conditions are met:

\begin{itemize}
 \item For some $\rho>0$, $f(\lambda)$ is analytic in a domain $D^k_\rho$.
\item The function $f(\lambda)$ admits $\sum_{n=0}^\infty a_n \lambda^n$ as a strong 
asymptotic expansion to all orders as $\lambda$ $\rightarrow 0$ with uniform estimate in $D^k_\rho$:
\begin{equation}  \label{taylorrem}
\left|  R^N f  \right|\leqslant A B^N [(kN)!]\lambda|^{N}.
\end{equation}
where $A$ and $B$ are some constants.
\end{itemize}
Then the Borel-Leroy transform of order $k$ defined by
\begin{equation}
B^{(k)}_f(u)=\sum_{n=0}^\infty \frac{a_n}{ (kn)!}u^n,
\end{equation}
is holomorphic for $|u|<B^{-1}$, it admits (for some $R>0$) an analytic continuation to the strip 
$\{u\in \mathbb C: |\Im u|< R, \Re u>0\}$ which does not grow too fast at infinity, so that  one recovers 
$f(\lambda)$ for $\lambda \in D^k_\rho$ through the absolutely convergent inverse integral
\begin{equation}\label{borinver}
 f(\lambda)=\frac{1}{k\lambda}\int_{0}^{\infty}B^{(k)}_f (u) e^{-(\frac{u}{\lambda})^{\frac{1}{k}}}\bigl(\frac{u}{\lambda}\bigr)^{\frac{1}{k}-1} du .
\end{equation}

\end{theorem}
\prf For $k=1$ this is exactly Nevanlinna's theorem as redicovered by Sokal \cite{Sok}. For larger values of $k$,
defining $g=\lambda^{1/k}$, we see that $\tilde f (g) =f(\lambda) \simeq \sum_n a_n g^{kn}$ satisfies Nevanlinna's hypothesis, and 
through the change of variables $u = \lambda v^k$ \eqref{borinver} is nothing but the ordinary inverse Borel formula 
from $\tilde B $ to $\tilde f$. \qed

\subsection{$\phi^{2k}$ theory in 0 dimension}
\label{borelsub}
The partition function of the $\phi^{2k}$ scalar theory in zero dimension for $k \ge 2$ is given by the one-dimensional integral
\begin{equation}\label{stand1}
Z_k(\lambda)= \int_{-\infty}^{+\infty} \frac{d\phi}{\sqrt{2\pi}} e^{-\frac{1}{2}\phi^2} e^{-\lambda\phi^{2k}/2} = 
\int d \mu (\phi) e^{-\lambda\phi^{2k}/2} ,
\end{equation}
where $d \mu$ is the normalized one-dimensional Gaussian measure of covariance 1. Its ``free energy" is simply $\log Z_k( \lambda)$.
The integral \eqref{stand1} will from now on be called the \emph{standard representation} of the theory.
The factor $1/2$ in front of $\lambda$ is a suitable normalization to simplify  the intermediate field representation below.\\

Illegally commuting series and integration leads to 
\be\label{2kpert}
Z_k(\lambda) \simeq \sum_{n=0}^{\infty}  a_{k,n}  \lambda^n , \ 
a_{k,n} = \frac{ (-1)^n }{2^n n!} \int d \mu (\phi) \phi^{2kn} =  (-1)^n\frac{(2kn)!! }{2^n n!}
\ee
where we define $2p !! =\prod_{k=1}^p (2k-1)$. Of course the power series $a_n$ has zero radius of convergence but it is Borel-Leroy summable 
of order $k-1$ (see Theorem \ref{mainbor} below).

The model exists also in a complex version, with partition function
\begin{equation} \label{compkmodel}
Z^c_{k}(\lambda)= \int_{-\infty}^{+\infty}\int_{-\infty}^{+\infty} \frac{d\phi d \bar \phi}{\pi} e^{-\bar\phi \phi} e^{-\lambda(\bar\phi \phi)^{k}} = 
\int_{\mathbb C} d \mu^c (\phi)e^{-\lambda(\bar\phi \phi)^{k}} ,
\end{equation}
where $d \mu^c$ is the normalized one-dimensional complex Gaussian measure of covariance 1. Its perturbative expansion is
\be\label{2kcpert}
Z^c_{k}(\lambda) \simeq \sum_{n=0}^{\infty}  a^c_{k,n}  \lambda^n , \
a_{k,n} = \frac{ (-1)^n }{2^n n!} \int d \mu^c (\phi) (\bar\phi \phi)^{kn} = (-1)^n \frac{(kn)!  }{n!}
\ee

\begin{theorem}\label{mainbor}
The partition functions $Z_k(\lambda)$ and $Z_{k,c}(\lambda)$  
are  Borel-Leroy summable of order $k-1$.\label{th1}
\end{theorem}
\prf Let us give the proof only for $Z_k (\lambda)$  as the complex case is similar. We shall prove analyticity in a domain 
$D^k_\rho=\{\lambda\in {\cal L}: \Re \lambda^{-1/k}> \rho^{-1}\}$ obviously bigger than the needed domain $D^{k-1}_\rho$.
But we shall prove Borel-Leroy summability of order $k-1$ (not k) since we shall prove Taylor remainder estimates in $[(k-1)N]!$ (see \eqref{2kpert}-\eqref{2kcpert}).

The integrand of $Z_k (\lambda)$ is an entire function of $\lambda$ which is uniformly bounded
by the integrable function $e^{-\frac{1}{2}\phi^2} $ in the right-half complex plane ${\mathbb C}_+ = \{ \lambda \, \vert \,\Re  (\lambda) >0\}$. 
Hence $Z_k$ is analytic in ${\mathbb C}_+$. For $k=2$, ${\mathbb C}_+$ contains
a disk $D^1_\rho$. For $k>2$ we continue $Z_k (\lambda)$ analytically to a domain of ${\cal L}$ of wider opening angle
by performing the change of variable $\phi = \lambda^{-\frac{1}{2k}} \psi$. Rotating the integration contour we find
\be  Z_k (\lambda)  =\lambda^{-\frac{1}{2k}}
\int_{- \infty}^{+ \infty}  e^{- \frac{\psi^{2k}}{2}}  e^{-\lambda^{-\frac{1}{k}} \frac{ \psi^{2}}{2}} \frac{d \psi}{\sqrt{2\pi}}.
\ee
In the domain $D^k_\rho$ the integrand is analytic and its absolute value is uniformly bounded
by the integrable function $ e^{- \frac{\psi^{2k}}{2}}$, hence we can conclude to analyticity of the integral $ Z_k (\lambda)$ in this domain.\footnote{We could without too much pain prove analyticity in a domain of larger opening angle, but shall not need it.}
However because of the prefactor $\lambda^{-\frac{1}{2k}}$ it is not yet obvious that $Z_k (\lambda)$ is uniformly bounded in 
$D^k_\rho$, as it should for \eqref{taylorrem} to hold at $N=0$. This can be checked through a single expansion step on the
$ e^{- \frac{\psi^{2k}}{2}} $ factor, writing 
\be e^{- \frac{\psi^{2k}}{2}} = 1 -\int_0^1 dt  \frac{\psi^{2k}}{2}e^{- t\frac{\psi^{2k}}{2}} ,\ee  
taking one $\psi$ factor out of $\psi^{2k}$ and joining it to  the Gaussian factor $e^{-\lambda^{-\frac{1}{k}} \frac{ \psi^{2}}{2}}$ to create a
full derivative (correcting for the missing $-\lambda^{-\frac{1}{k}}$ factor) and then 
performing integration by parts. It leads to:
\begin{eqnarray} \label{integconsta}
 Z_k (\lambda)  &=&   1 -   \lambda^{-\frac{1}{2k}}
\int_{- \infty}^{+ \infty} \frac{d \psi}{2\sqrt{2\pi}} \int_0^1 dt  e^{- t\frac{\psi^{2k}}{2}} \psi^{2k-1} \biggl[ -\lambda^{\frac{1}{k}}\frac{d}{d\psi }
e^{-\lambda^{-\frac{1}{k}} \frac{ \psi^{2}}{2}}\biggr]
\\ \nonumber
&=&1 -  \lambda^{\frac{1}{2k}}
\int_{- \infty}^{+ \infty} \frac{d \psi}{2\sqrt{2\pi}} \int_0^1 dt [2k-1-  t k\psi^{2k}]\psi^{2k-2} e^{- t\frac{\psi^{2k}}{2}-\lambda^{-\frac{1}{k}} \frac{ \psi^{2}}{2}},
\end{eqnarray}
an expression now easy to bound for $\lambda \in D^k_\rho$ by a ($k$ dependent) constant.

The uniform estimates in $A_k B_k^N [(kN)!] \vert \lambda|^{N}$ of \eqref{taylorrem} for $N \ge 1$ (the constants $A_k$ 
and $B_k$ being of course allowed to depend on $k$), are similar. First:
\begin{eqnarray}\label{restetay}
 R^N Z_{k}(\lambda) &=&  \lambda^{N}
\int_0^1 \frac{(1-t)^{N-1}}{(N-1)!} dt \int d \mu (\phi)  (-\frac{\phi^{2k}}{2})^N  e^{-t\lambda\frac{\phi^{2k}}{2}}   \\
&= &  \frac{ (-1)^N\lambda^{-\frac{1}{2k}}}{2^{N}(N-1)!} \int_0^1 (1-t)^{N-1} dt  \int_{- \infty}^{+ \infty} \frac{d \psi}{\sqrt{2\pi}}   \psi^{2kN} 
  e^{-t \frac{\psi^{2k}}{2}-\lambda^{-\frac{1}{k}} \frac{ \psi^{2}}{2}} .\nonumber 
\end{eqnarray}
Let us evaluate 
\be
I_N = \frac{1}{(N-1)!}  \int_{- \infty}^{+ \infty} \frac{d \psi}{\sqrt{2\pi}}   \psi^{2kN} 
  e^{-t \frac{\psi^{2k}}{2}-\lambda^{-\frac{1}{k}} \frac{ \psi^{2}}{2}} .
\ee
Applying as before a single expansion step on the
\be e^{- t\frac{\psi^{2k}}{2}} = 1 -t\int_0^1 dt'  \frac{\psi^{2k}}{2}e^{- tt'\frac{\psi^{2k}}{2}} 
\ee factor and computing exactly the first factor which is a Gaussian integral, we find
$I_N =  \frac{ (2kN)!!}{(N-1)!} \lambda^{N+\frac{1}{2k}} - tJ_N$, with
\be
J_N := \frac{1}{(N-1)!}  \int_0^1 dt'  \int_{- \infty}^{+ \infty} \frac{d \psi}{2\sqrt{2\pi}}   \psi^{2k(N+1)} 
  e^{-t t'\frac{\psi^{2k}}{2}-\lambda^{-\frac{1}{k}} \frac{ \psi^{2}}{2}}.
\ee
To bound $J_N$, we consider the factor $ \psi^{2k(N+1)} $ as an initial vertex, of coordination $2k(N+1)$.

We apply exactly $kN+1$ Wick contraction steps to (less than half) the fields of this initial vertex 
with respect to the Gaussian measure $e^{-\lambda^{-\frac{1}{k}} \frac{ \psi^{2}}{2}} $. 
Each such step is  similar to the one of \eqref{integconsta}. More precisely each step
\begin{itemize}

\item selects a remaining field in the initial vertex $ \psi^{2k(N+1)} $, join it to  the Gaussian factor $e^{-\lambda^{-\frac{1}{k}} \frac{ \psi^{2}}{2}}$ to create a full derivative (correcting for the missing $-\lambda^{-\frac{1}{k}}$ factor)

\item then integrates by parts: the derivative either acts on the interaction $ e^{- t\frac{\psi^{2k}}{2}}$ or on the remaining fields 
of the initial vertex or on the field created by previous integration steps.

\end{itemize}

The process cannot run short of fields in the initial vertex $ \psi^{2k(N+1)} $ because each step consumes at most two of these fields and the number of steps is $kN+1$, less than \emph{half} the total number $(2kN+2k)$ of such fields. We call $p$ the number of times the integration by parts hits the exponential $e^{-t t'\frac{\psi^{2k}}{2}}$; hence $0\le p \le kN+1 $, and it hits $ kN +1 - p$ times the fields down from the exponential.
The result is a complicated sum of non perturbative amplitudes $A_G$ for processes $G$\footnote{Processes are not exactly Feynman graphs 
because of the remaining non-perturbative interaction factor $ e^{-tt' \frac{\psi^{2k}}{2}}$.}. They all have exactly $kN+1$ edges, hence $kN+1$ covariance factors $\lambda^{1/k}$, which in total give a factor $\lambda^{N+\frac{1}{k}}$. The number of fields down at the end must be
$ 2k(N+1) + p(2k-2)  -2 (kN +1 - p) =  2k(p+1) -2$ because $p$ steps destroy a field and create $2k-1$ fields (together with a $tt'k$ factor)
and $kN +1 - p$ steps destroy two fields.

Therefore any process with a given value of $p$ has the same amplitude
\be J_{N,p} := \lambda^{N+\frac{1}{k}}  \int_0^1 dt'  (-tt'k)^p  \int_{- \infty}^{+ \infty} \frac{d \psi}{2\sqrt{2\pi}}  \psi^{2k(p+1) -2 } 
  e^{-tt' \frac{\psi^{2k}}{2}-\lambda^{-\frac{1}{k}} \frac{ \psi^{2}}{2}}.
\ee

\begin{lemma} 
\be  \vert J_{N,p} \vert \le \vert \lambda\vert^{N+\frac{1}{k}}  A_k B_k^N  p! \quad 
\ee
for some constants $A_k$ and $B_k$ (possibly depending on $k$).
\end{lemma}
\prf If $p=0$, $J_{N,0}$ is bounded by a ($k$ dependent) constant times $\vert \lambda\vert^{N+\frac{1}{k}} $ since 
 in $D^k_\rho$ we have $\Re\lambda^{-\frac{1}{k}}  > \rho^{-1}$. 
If $p\ge 1$,
\be J_{N,p} =   (-k)^{p}\lambda^{N+\frac{1}{k}} \int_0^1 dt'     \int_{- \infty}^{+ \infty} \frac{d \psi}{2\sqrt{2\pi}}  K_{N,p}(\psi,t,t')L (\psi),
\ee
where we define 
\bea K_{N,p}(\psi, t,t') &:=&   (tt')^p \psi^{2k(p-1)+k} e^{-t t'\frac{\psi^{2k}}{2}} \\
L (\psi) &:=&  \psi^{3k-2}  e^{-\lambda^{-\frac{1}{k}} \frac{ \psi^{2}}{2}}.
\eea
Applying a Cauchy-Schwarz bound we find
\be \vert J_{N,p} \vert \le  k^p\vert \lambda\vert^{N+\frac{1}{k}}  \int_0^1 dt'   \sqrt{\int_{- \infty}^{+ \infty} \frac{d \psi}{2\sqrt{2\pi}}   K^2_{N,p}(\psi, t,t')\int_{- \infty}^{+ \infty} \frac{d \psi}{2\sqrt{2\pi}}   L^2(\psi)} .
\ee
Obviously the integral over $L$ is easily bounded by a $k$ dependent constant since 
 in $D^k_\rho$ we have $\Re\lambda^{-\frac{1}{k}}  > \rho^{-1}$. The other integral is also easily bounded by the change
of variables $u = t t'\psi^{2k}$. Indeed
\bea \sqrt{\int_{- \infty}^{+ \infty} \frac{d \psi}{2\sqrt{2\pi}}   K^2_{N,p}(\psi, t,t')} &=& \sqrt{\int_{- \infty}^{+ \infty} \frac{d \psi}{2\sqrt{2\pi}}
   (tt')^{2p} \psi^{4k(p-1)+2k}e^{-t t'\psi^{2k}}}\nonumber\\
&\le& \sqrt{\frac{(tt')^{\frac{2k-1}{2k}}}{2k\sqrt{2\pi}}\int_0^\infty u^{2p-2+\frac{1}{2k}} e^{-u} du}\nonumber\\
&\le& A_k B_k^N p!
\eea
since $tt' \le 1$. Remembering that $p\le kN +1$ completes the proof.\qed

To complete the bound on $J_N$ we need only to multiply this bound on $J_{N,p}$ by $\frac{1}{(N-1)!} $ times the number of processes at given $p$, then sum over $p$.
But it is not necessary to compute the exact number of such processes with a fixed $p$. We can just give a crude bound on it (we do not try to find optimal constants). This number is bounded by $2^{kN+1}$ (to choose the $p$ particular steps which derive the exponential)
times the product over the
$ kN +1 - p$ steps which derive fields down from the exponential of the number of fields down
the exponential at that step. This last number is certainly at most
the total maximal number of fields ever produced down the exponential, raised to the power $kN+1-p$, 
hence certainly bounded by
\be  [2k(N+1) + (2k-2)p]^{kN+1-p} \le N^{kN+1-p}  [2k^2 +4k]^{kN+1}.
\ee
Taking into account that $N^{1-p}p!  \le N(k+1)^{(k+1)N}$ for $p\le kN+1$, we get a bound of the form $ A_k B_k^N N^{kN}$,
which is independent of $p$. Summing over $p$ adds a factor $kN +1$ which can be absorbed by changing $B_k$. Finally
multipying by $\frac{1}{(N-1)!} $ (and using Stirling's formula) gives a bound of the form $ A_k B_k^N N^{(k-1)N}$.
Returning to \eqref{restetay} 
and gathering all factors proves  \eqref{taylorrem} hence completes the proof of Theorem \ref{mainbor}. \qed

From now on our goal is to define new intermediate field representations for the
functions $Z_k(\lambda)$, $Z_{k,c}(\lambda)$  
and to check in these new representations their Borel-Leroy summability.

\section{Imaginary Gaussian IF Representation}
\label{heuristic}

\subsection{Real Case}
\medskip

We first split the interaction in two 
using an intermediate field $\sigma$ with normalized Gaussian measure $d\mu (\sigma)$
of covariance 1. The result is:
\begin{equation}
e^{- \lambda\phi^{2k}/2}=\int{d\mu (\sigma)}e^{i\sqrt{\lambda}\phi^k\sigma}.
\end{equation}
We define $g_k = \lambda^{\frac{1}{2k}}$, and as next step we decompose
\begin{equation}  \label{genek}
i\sqrt{\lambda}\phi^k \sigma=\frac{i}{4}[(g_k \phi
\sigma+(g_k\phi)^{k-1})^2-(g_k\phi\sigma-(g_k\phi)^{k-1})^2 ] .
\end{equation}
We introduce a pair of intermediate fields $a_1$ and $b_1$  with imaginary covariances $-i$ and $+i$, hence the Gaussian measure
$d\mu_{\pm i}(a_1, b_1) = d\mu_{-i} (a_1) d\mu_{i} (b_1) $ 
 so that 
\begin{eqnarray}
e^{i\sqrt{\lambda}\phi^k\sigma}&=&\int d\mu_{\pm i}(a_1, b_1)  
e^{\frac{i}{\sqrt 2} [(g_k\phi \sigma + (g_k\phi)^{k-1})a_1 + (g_k\phi \sigma - (g_k\phi)^{k-1})b_1 ]}\quad \\
&=&\int d\mu_{\pm i}(a_1, b_1)   e^{i [  g_k\phi\sigma \frac{a_1 +b_1}{\sqrt 2} + (g_k\phi)^{k-1}\frac{a_1 -b_1}{\sqrt 2} ]}.
\end{eqnarray}

We now change variables for 
\begin{eqnarray}
\label{eqref:ChdeVr}
\alpha_1=\frac{a_1+b_1}{\sqrt2} , \qquad\qquad \beta_1=\frac{a_1-b_1}{\sqrt2},
\end{eqnarray}
so that 
\begin{eqnarray}
e^{i\sqrt{\lambda}\phi^k\sigma}&=&\int d\mu_X(\alpha_1, \beta_1)   e^{i g_k\phi\sigma \alpha_1 +  i (g_k\phi)^{k-1}\beta_1 },
\end{eqnarray}
where the Gaussian measure $d\mu_X$ is defined by its covariance
\begin{eqnarray}
\label{eqref:CrossMeas1}
 <\alpha_1\beta_1>_X=-i,\  \ \quad<\alpha_1^2>_X=0,\  \ \quad<\beta_1^2>_X=0 .
\end{eqnarray}
Remember this section is heuristic so do not worry yet about convergence and integration contours,
which will be addressed in the next section.

We keep the term $i g_k\phi \sigma  \alpha_1$ 
and decompose the $i g_k^{k-1}\phi^{k-1}\beta_1 $ term as 
\begin{equation}  \label{genek-1}
e^{ig_k^{k-1}\phi^{k-1}\beta_1 }=\int d\mu_X(\alpha_2, \beta_2)  e^{ig_k\phi
\beta_1  \alpha_2 +i (g_k\phi)^{k-2}\beta_2 }.
\end{equation}

Continuing in this way we prove inductively the following representation:
\begin{equation}
e^{- \lambda\phi^{2k}/2 }  =\int d\mu (\sigma)  \prod_{j=1}^{k-1} d\mu_X(\alpha_j, \beta_j) e^{i g_k [\phi \sigma  \alpha_1 +   \sum_{j=1}^{k-2} \phi \beta_j   \alpha_{j+1}  + \phi \beta_{k-1} ]}, 
\end{equation}
where the $\alpha_j$, $\beta_j$ and the measure $d\mu_X$ are respectively defined as in (\ref{eqref:ChdeVr})  and (\ref{eqref:CrossMeas1}).

We now integrate 
%
\begin{itemize}
\item for $k$ odd, over $\phi$, $\sigma$ and all even $\alpha_{2j}$, $\beta_{2j}$, for $j\in\{1,..,\frac{k-1}{2}\}$. In that case we denote
$\Phi=(\phi, \sigma, \alpha_2,\beta_2 ,...    , \alpha_{k-1}, \beta_{k-1})$ the $k+1$ integrated variables and 
$\Psi=(\alpha_1,\beta_1,..., \alpha_{k-2}, \beta_{k-2})$ the $k-1$ remaining ones. The Gaussian measure
$d\mu (\sigma)  \prod_{j=1}^{k-2} d\mu_X(\alpha_j, \beta_j)$
factorizes as $d\nu(\Phi) d\chi (\Psi) $.

\item for $k$ even, over $\phi$ and all odd $\alpha_{2j-1}$, $\beta_{2j-1}$, for $j\in\{1,..,\frac{k}{2}\}$.  In that case we denote
$\Phi=(\phi,  \alpha_1,\beta_1 ,...    , \alpha_{k-1}, \beta_{k-1})$ the $k+1$ integrated variables and 
$\Psi=(\sigma, \alpha_2,\beta_2,..., \alpha_{k-2} , \beta_{k-2})$ the $k-1$ remaining ones. The Gaussian measure
$d\mu (\sigma)  \prod_{j=1}^{k-2} d\mu_X(\alpha_j, \beta_j)$
factorizes again as $d\nu(\Phi) d\chi (\Psi) $.
\end{itemize}

The partition function writes 
\begin{equation} \label{interrepreal}
Z_k(\lambda)=\int d\chi(\Psi)  \biggl[d\nu (\Phi)\exp[\frac{ig_k}{2} < \Phi, H_k(\Psi).\Phi >] \biggr], 
\end{equation}
where $H_k$ is a   $(k+1)\times (k+1)$ real symmetric matrix. More precisely :
%
\begin{itemize}
\item if $k=2p+1$ is odd,   $H_k$ is
\begin{equation}
H_{k}= \left(
\begin{array}{c|ccccc}
  0 &  \alpha_1 & \beta_1 &  \cdots & \beta_{k-2}  & 1 \\ \hline
  \alpha_1 & &&\raisebox{-30pt}{{\huge\mbox{{$0$}}}} \\[-6ex]
  \beta_1& \\[-0.5ex]
  \vdots & \\[-0.5ex]
  \beta_{k-2}& \\
1& \\
[-0.5ex]  
\end{array}
\right), \hspace{0.7cm}
\end{equation}
\item if $k=2p$ is even,  $H_k$ is 
\begin{equation}
H_{k}= \left(
\begin{array}{c|ccccc}
  0 &  \sigma & \alpha_2  &  \cdots & \beta_{k-2} & 1  \\ \hline
   \sigma & &&\raisebox{-30pt}{{\huge\mbox{{$0$}}}} \\[-6ex]
  \alpha_2& \\[-0.5ex]
  \vdots & \\[-0.5ex]
  \beta_{k-2}& \\
 1& \\[-0.5ex]  
\end{array}
\right) . \hspace{1.35cm}
\end{equation}
\end{itemize}
The Gaussian integration over $\Phi$ gives a determinant. Rewritten as usual in field theory as 
exponential of an action, it leads to 
the \emph{IF representation}
\label{theorintrep}
\begin{equation}
Z_k(\lambda)=\int d\chi(\Psi)\exp\bigl[-\frac{1}{2}\Tr\ln(1-g_k M_k(\Psi))\bigr],  \label{intfieldrep}
\end{equation} 
where $M_k(\Psi)=iC_k.H_k(\Psi)$ and $C_k$, the covariance for the $\Phi$ variables, is
\begin{eqnarray}
\label{eqref=Codd}
\newcommand*{\temp}{\multicolumn{1}{c|}{}}
\newcommand*{\tempi}{\multicolumn{1}{c|}{-i}}
\newcommand*{\tempo}{\multicolumn{1}{c|}{0}}
C_{odd}=\left(\begin{array}{ccccccc}
1&\multicolumn{1}{c|}{0} &&&&& \\ 
0&\multicolumn{1}{c|}{1} &&&&& \\ 
\cline{1-4}
& \temp & 0 &  \tempi &  &\Large\mbox{{$0$}} & \\ 
& \temp & - i & \tempo &  & & \\ 
\cline{3-6} 
&&& \temp & 0 & \tempi & \\
& &\Large\mbox{{$0$}}& \temp & -i & \tempo & \\
\cline{5-6} 
&&&&&&\ddots
\end{array}\right) ,
\end{eqnarray}
\begin{eqnarray}
\label{eqref=Ceven}
\newcommand*{\temp}{\multicolumn{1}{c|}{}}
\newcommand*{\tempi}{\multicolumn{1}{c|}{-i}}
\newcommand*{\tempo}{\multicolumn{1}{c|}{0}}
C_{even}=\left(\begin{array}{cccccc}
\multicolumn{1}{c|}{1} &&&&& \\ 
\cline{1-3}
 \temp & 0 &  \tempi &  &\Large\mbox{{$0$}} & \\ 
 \temp & - i & \tempo &  & & \\ 
\cline{2-5} 
&& \temp & 0 & \tempi & \\
 &\Large\mbox{{$0$}}& \temp & -i & \tempo & \\
\cline{4-5} 
&&&&&\ddots
\end{array}\right).
\end{eqnarray}
\vspace{0.3cm}

The proof that this integral representation \eqref{intfieldrep} converges and that the integral is indeed $Z_k(\lambda)$ is postponed to Section III. 
In the simplest cases $k=3,4$, hence for the $e^{-\lambda\phi^{6}/2}$ and  $e^{-\lambda\phi^{8}/2}$ models, we obtain the representations:

\begin{equation} Z_3( \lambda) =\int d\chi (\alpha_1, \beta_1) e^{- \frac{1}{2}\Tr \ln  [1 - \lambda^{1/6} M_3]}, \ 
M_3= \begin{pmatrix}
0  &  i\alpha_1 &i\beta_1 & i \\
i\alpha_1 & 0 & 0 &0 \\
1 & 0 & 0 &0 \\
\beta_1 & 0 & 0 & 0
\end{pmatrix}  
\ee
and
\begin{equation} Z_4( \lambda) = \int d\chi (\sigma, \alpha_2, \beta_2) e^{- \frac{1}{2}\Tr \ln  [1 - \lambda^{1/8} M_4]}, \
M_4= \begin{pmatrix}
0  & i\sigma & i\alpha_2 & i\beta_2 & i \\
\alpha_2 & 0 & 0 & 0 & 0 \\
\sigma & 0 & 0 & 0 & 0 \\
1 & 0 & 0 & 0 & 0 \\
\beta_2 & 0 & 0 & 0 & 0
\end{pmatrix}  .
\ee
\subsection{Complex Case}
\medskip

As in the previous section, we first split the interaction in two 
using a complex intermediate field $\sigma$ with normalized Gaussian measure $d\mu (\sigma)$
of covariance 1. We obtain
\begin{eqnarray}
&e^{- \lambda(\phi\bar \phi)^{2p+1}}&=\int{d\mu^c (\sigma)}e^{i\sqrt{\lambda}(\phi\bar\phi)^p(\bar\phi\sigma+\phi\bar\sigma)}, \\
&e^{- \lambda(\phi\bar \phi)^{2p}}&=\int{d\mu^c (\sigma)}e^{i\sqrt{\lambda}(\phi\bar\phi)^p(\sigma+\bar\sigma)}.
\end{eqnarray}
We define $g_k = \lambda^{\frac{1}{2k}}$, and as next step we decompose
\begin{eqnarray}  \label{genekc}
i\sqrt{\lambda}(\phi\bar\phi)^p (\bar\phi\sigma+\phi\bar\sigma)&=&\frac{i}{2}[\lvert g_k \bar\phi
\sigma+(g_k^2\phi\bar\phi)^p\rvert^2-\lvert g_k\bar\phi\sigma-(g_k^2\phi\bar\phi)^{p}\rvert^2 ],\\
i\sqrt{\lambda}(\phi\bar\phi)^p (\sigma+\bar\sigma)&=&\frac{i}{2}[\lvert g_k \bar\phi
\sigma+(g_k^2\phi\bar\phi)^{p-1}g_k\bar\phi\rvert^2-\lvert g_k\bar\phi\sigma-(g_k^2\phi\bar\phi)^{p-1}g_k\bar\phi\rvert^2 ].\nonumber
\end{eqnarray}
We introduce a pair of complex intermediate fields $a_1$ and $b_1$  with imaginary covariances $-i$ and $+i$, hence the Gaussian measure
\be d\mu^c_{\pm i}(a_1, b_1) = d\mu^c_{-i} (a_1,\bar a_1) d\mu^c_{i} (b_1, \bar b_1) .
\ee
More precisely it means that writing $a_1^R, b_1^R$ and $a_1^I, b_1^I$ for the real and imaginary parts of $a_1$ and $b_1$, the measure 
$d\mu^c_{-i} (a_1,\bar a_1) d\mu^c_{i} (b_1, \bar b_1)$ is the product of four independent Gaussain measures on these four real variables.
Each of the four can be independently complexified and 
we should use an appropriate contour of integration $C_{+,\epsilon}$ in each of the four corresponding complex planes, as defined in subsection \ref{imag}. However remember that in this heuristic section we do not consider convergence questions.
Then we have
\begin{eqnarray}
e^{i\sqrt{\lambda}(\phi\bar\phi)^p (\bar\phi\sigma+\phi\bar\sigma)}&=&\int d\mu^c_{\pm i}(a_1, b_1)  
e^{\frac{i}{\sqrt 2} [(g_k \bar\phi\sigma+(g_k^2\phi\bar\phi)^p)a_1 } \\
&&\hspace{3.2cm}^{+ (g_k\bar\phi\sigma-(g_k^2\phi\bar\phi)^{p})b_1 + c.c.]}\nonumber\\
&=&\int d\mu^c_{\pm i}(a_1, b_1)   e^{i [  g_k\bar \phi\sigma \frac{a_1 +b_1}{\sqrt 2} + (g_k^2\phi\bar\phi)^{p}\frac{a_1 -b_1}{\sqrt 2} + c.c.]},
\nonumber\\
\nonumber\\
e^{i\sqrt{\lambda}(\phi\bar\phi)^p (\sigma+\bar\sigma)}&=&\int d\mu^c_{\pm i}(a_1, b_1)   e^{i [  g_k\bar \phi\sigma \frac{a_1 +b_1}{\sqrt 2} + (g_k^2\phi\bar\phi)^{p}g_k\bar\phi\frac{a_1 -b_1}{\sqrt 2} + c.c.]},\nonumber
\end{eqnarray}
where $c.c.$ means complex conjugate. We can change variables as in the real case for
\begin{eqnarray}\label{eqref:ChdeVrc}
\alpha_1=\frac{a_1+b_1}{\sqrt2}, \qquad\qquad \beta_1=\frac{a_1-b_1}{\sqrt2},
\end{eqnarray}
and complex conjugates, so that 
\begin{eqnarray}
e^{i\sqrt{\lambda}(\phi\bar\phi)^p (\bar\phi\sigma+\phi\bar\sigma)}&=&\int d\mu^c_X(\alpha_1, \beta_1)   e^{i [  g_k\bar \phi\sigma \alpha_1 + (g_k^2\phi\bar\phi)^{p}\beta_1 + c.c.]},\\
e^{i\sqrt{\lambda}(\phi\bar\phi)^p (\sigma+\bar\sigma)}&=&\int d\mu^c_X(\alpha_1, \beta_1)   e^{i [  g_k\bar \phi\sigma\alpha_1 + (g_k^2\phi\bar\phi)^{p}g_k\bar\phi\beta_1 + c.c.]},
\end{eqnarray}
where the Gaussian measure $d\mu^c_X(\alpha_1, \beta_1)$ is defined by its covariance   
\begin{eqnarray}
\label{eqref:CrossMeasc}
 &&<\alpha_1\bar\beta_1>_X= <\bar\alpha_1\beta_1>_X=-i,  \label{surp} \\
&&<\alpha_1^2>_X=<\bar\alpha_1^2>_X=<\beta_1^2>_X=<\bar\beta_1^2>_X=0
\\   
&& <\alpha_1\bar \alpha_1 >_X= <\beta_1 \bar \beta_1>_X= <\alpha_1\beta_1 >_X= <\bar \alpha_1 \bar \beta_1>_X =0.
\end{eqnarray}
We invite the reader to check in particular \eqref{surp}, which may be surprising at first sight but is perfectly consistent
with the imaginary-Gaussian complex integration rule \eqref{imgau}).

An inductive reasoning strictly parallel to the previous subsection leads to:
\begin{equation}
e^{- \lambda(\phi\bar\phi)^{k} }  =\int d\mu^c (\sigma)  \prod_{j=1}^{k-1} d\mu^c_X(\alpha_j, \beta_j) e^{i g_k [\bar\phi \sigma  \alpha_1 +   \sum_{j=1}^{k-2} \bar\phi \beta_j   \alpha_{j+1}  + \bar\phi \beta_{k-1} + c.c. ]}, 
\end{equation}
where the $\alpha_j$, $\beta_j$ and the measure $d\mu^c_X$ are respectively defined as in (\ref{eqref:ChdeVrc})  and (\ref{eqref:CrossMeasc}).
Integrating again
\begin{itemize}
\item for $k$ odd, over $\phi$, $\sigma$, all even $\alpha_{2j}$, $\beta_{2j}$, for $j\in\{1,..,\frac{k-1}{2}\}$ and complex conjugates. In that case we denote
$\Phi=(\phi, \sigma, \alpha_2,\beta_2 ,...    , \alpha_{k-1}, \beta_{k-1})$ the $k+1$ integrated variables and 
$\Psi=(\alpha_1,\beta_1,..., \alpha_{k-2}, \beta_{k-2})$ the $k-1$ remaining ones. The Gaussian measure
$d\mu^c (\sigma)  \prod_{j=1}^{k-2} d\mu^c_X(\alpha_j, \beta_j)$
factorizes as $d\chi(\Psi,\bar\Psi) d\nu (\Phi,\bar\Phi)$.

\item for $k$ even, over $\phi$, all odd $\alpha_{2j-1}$, $\beta_{2j-1}$, for $j\in\{1,..,\frac{k}{2}\}$ and complex conjugates.  In that case we denote
$\Phi=(\phi,  \alpha_1,\beta_1 ,...    , \alpha_{k-1}, \beta_{k-1})$ the $k+1$ integrated variables and 
$\Psi=(\sigma, \alpha_2,\beta_2,..., \alpha_{k-2} , \beta_{k-2})$ the $k-1$ remaining ones. The Gaussian measure
$d\mu^c (\sigma)  \prod_{j=1}^{k-2} d\mu^c_X(\alpha_j, \beta_j)$
factorizes again as $d\chi(\Psi,\bar\Psi) d\nu (\Phi,\bar\Phi) $.
\end{itemize}
the partition function writes 
\begin{equation}
Z_{k,c}(\lambda)=\int d\chi(\Psi,\bar\Psi)  \biggl[d\nu (\Phi,\bar\Phi)\exp[ ig_k < \Phi, H_k(\Psi, \bar\Psi).\Phi >] \biggr], 
\end{equation}
where $H_k$ is a   $(k+1)\times (k+1)$ Hermitian matrix. More precisely :
\begin{itemize}

\item if $k=2p+1$ is odd,   $H_k$ is
\begin{equation}
H_{k}= \left(
\begin{array}{c|ccccc}
  0 &  \alpha_1 & \beta_1 &  \cdots & \beta_{k-2}  & 1 \\ \hline
  \bar\alpha_1 & &&\raisebox{-30pt}{{\huge\mbox{{$0$}}}} \\[-6ex]
  \bar\beta_1& \\[-0.5ex]
  \vdots & \\[-0.5ex]
  \bar\beta_{k-2}& \\
1& \\
[-0.5ex]  
\end{array}
\right), \hspace{0.7cm}
\end{equation}
\item if $k=2p$ is even,  $H_k$ is 
\begin{equation}
H_{k}= \left(
\begin{array}{c|ccccc}
  0 &  \sigma & \alpha_2  &  \cdots & \beta_{k-2} & 1  \\ \hline
  \bar \sigma & &&\raisebox{-30pt}{{\huge\mbox{{$0$}}}} \\[-6ex]
 \bar \alpha_2& \\[-0.5ex]
  \vdots & \\[-0.5ex]
 \bar \beta_{k-2}& \\
 1& \\[-0.5ex]  
\end{array}
\right) . \hspace{1.35cm}
\end{equation}
\end{itemize}
\vspace{0.3cm}

Gaussian integration over $\Phi$ leads to the
\emph{IF representation}
\begin{equation}\label{intfieldrepcc}
Z_{k,c}(\lambda)=\int d\chi(\Psi)\exp\bigl[-\Tr\ln(1-g_k M_k(\Psi))\bigr],
\end{equation} 
where $M_k(\Psi)=iC_k.H_k(\Psi)$ and $C_k$, the covariance for the $\Phi$ variables, is given as in the real case by (\ref{eqref=Codd}) and (\ref{eqref=Ceven}). \\

In the simplest cases $k=3,4$
we obtain the representations :
\begin{equation} Z_{3,c}( \lambda) =\int d\chi (\alpha_1, \beta_1) e^{- \Tr \ln  [1 - \lambda^{1/6} M_3]}, \ 
M_3= \begin{pmatrix}
0  &  i\alpha_1 &i\beta_1 &1  \\
i\bar\alpha_1 & 0 & 0 &0 \\
1 & 0 & 0 &0 \\
\bar\beta_1 & 0 & 0 & 0
\end{pmatrix}  
\ee
and
\begin{equation} Z_{4,c}( \lambda) = \int d\chi (\sigma, \alpha_2, \beta_2) e^{- \Tr \ln  [1 - \lambda^{1/8} M_4]}, \ 
M_4= \begin{pmatrix}
0  & i\sigma & i\alpha_2 & i\beta_2 & i \\
\bar\alpha_2 & 0 & 0 & 0 & 0 \\
\bar\sigma & 0 & 0 & 0 & 0 \\
1 & 0 & 0 & 0 & 0 \\
\bar\beta_2 & 0 & 0 & 0 & 0
\end{pmatrix}  .
\ee
\section{Analyticity Domains}
\label{analyt}

In this section we prove the following theorem:
\begin{theorem}\label{maintheo}
The integral $\int d\chi(\Psi)\exp\bigl[-\frac{1}{2}\Tr\ln(1-g_k M_k(\Psi))\bigr]$
is absolutely convergent in a domain $D^{k-1}_\rho=\{\lambda\in \mathbb C: 
\Re \lambda^{-\frac{1}{k-1}}> \rho^{-1}\}$ (for $\rho$ sufficiently small). It defines a Borel-Leroy summable
function of order$k-1$ in this domain, whose Taylor series at the origin is the same as the Taylor series of 
$Z_k(\lambda)$. Hence (by unicity of the Borel sum) \eqref{intfieldrep} and \eqref{intfieldrepcc} hold.
\end{theorem}

\prf We give the proof in the real $\phi^{2k}$ case, the argument for the complex case $(\bar \phi \phi )^k$ being essentially identical.
The key step 
is an upper bound on the norm of the resolvent $[1-g_k M(\Psi)]^{-1}$ in the Nevanlinna domain for
Borel-Leroy summability of order $k-1$. This bound must be uniform both in $\lambda$
in that domain and uniform in the intermediate fields along the contours 
associated to $d \chi$. 

Let us prove such a uniform bound in a slightly 
larger domain $E^{k-1}_\rho$ consisting of all $\lambda = \rho e^{i\theta} $ with $\rho$ small and
$\vert\theta \vert < \frac{ (k-1) \pi }{2}$ (hence in a half-disk for $\lambda^{\frac{1}{k-1}}$). Obviously it contains
the disk $D^{k-1}_{\rho/2}$ necessary for Nevanlinna's Theorem.
We need to compute the eigenvalues of the matrix $1-g_k M_k$, and to take into account the contours of integration.

\begin{lemma} For $\lambda \in E^{k-1}_1$ and $\Psi$ on the contours of integration $C_{\pm \epsilon}$
with $\epsilon= \frac{1}{4}  k^{-1/2}  \sin \frac{\pi}{4k}  $  we have
\begin{equation}
\Vert (1-g_kM_k)^{-1}\Vert \le 2 [ \sin \frac{\pi}{4k} ]^{-1}. \label{resobound}
\end{equation} \label{analytlemma}
where, from now on the notation $\Vert M \Vert$ means the \emph{operator norm} of $M$.
\end{lemma}

\prf
Returning to the parametrization of our contour integrals we recall that $a_j = \Re a_j - i \epsilon \tanh (\Re a_j )$
and $b_j = \Re b_j - i \epsilon \tanh (\Re b_j ) $, where $\Re$ is the real part. Hence remembering 
\eqref{eqref:ChdeVr}, and putting $\Psi = X + i Y$, where the vectors $X$ and $Y$ are real, each 
coefficient of $Y$ is bounded in absolute value by $\epsilon\sqrt 2$.
The matrix $M(\Psi)$ being linear in $\Psi$, we have $M_k(\Psi) = M_k(X) + i M_k(Y)$, and since each of the $2k$ non-zero coefficients of $M_k(Y)$ 
is bounded in absolute value by $\epsilon\sqrt 2$, we can bound its Hilbert-Schmidt norm $\Vert M_k (Y) \Vert_{2} $ by $2\epsilon \sqrt k$, hence
\be \Vert M_k(Y) \Vert  \le \Vert M_k (Y) \Vert_{2} = 2\epsilon \sqrt k \label{HS}
\ee
Now let us compute the eigenvalues of the matrix $1-g_k M_k (X)$.
It has eigenvalue $1$ with multiplicity $k-1$ and two non trivial eigenvalues,
\begin{equation}
x_{\pm}= 1\pm g_k  \sqrt{ R_k},
\end{equation}
where  $R_k$ is 
$- (\Re \alpha_1)^2 + i ( \Re \beta_1 \Re \alpha_3 + \cdots  + \Re \beta_{k-4} \Re \alpha_{k-2}  + \Re \beta_{k-2} ) $ if $k$ is odd and is
$ i (\sigma \Re \alpha_2 + \Re \beta_2 \Re \alpha_4  + \cdots  + \Re \beta_{k-4} \Re \alpha_{k-2}  + \Re  \beta_{k-2})  $
if $k$ is even.

If $R_k$ is not zero we can state something about the argument of $\pm \sqrt {R_k}$.
In the odd case, if $\Re \alpha_1  \ne 0$, we have $R_k = -a^2 (1 + i b)$ with $a$ and $b$ real, hence 
$\pm \sqrt{ R_k} = i a \sqrt{1 + i b} $ and the argument of $ \pm \sqrt{ R_k} $ lies  in
$ I=  [\frac{ \pi} {4} , \frac{3 \pi} {4} ] \cup [-\frac{3 \pi} {4} ,  -\frac{ \pi} {4}  ] $.
If $\Re \alpha_1  = 0$ or  in the even case the argument of $\pm \sqrt{ R_k}$ belongs to $\{ -\frac{3 \pi} {4} ,  -\frac{ \pi} {4} , \frac{ \pi} {4} , 
\frac{3 \pi} {4}\}$, hence to the boundary of $I$.

But in the domain $E^{k-1}_\rho$
the argument of $g_k$  is bounded by $\frac{ (k-1) \pi }{4k}$ hence the argument of $ \pm g_k  \sqrt{ R_k} $ (when $g_k R_k \ne 0$) lies in 
\bea I_k & =&[\frac{ \pi} {4} -  \frac{ (k-1) \pi }{4k}, \frac{3 \pi} {4} +  \frac{ (k-1) \pi }{4k} ] \cup [-\frac{3 \pi} {4}-  \frac{ (k-1) \pi }{4k} ,  -\frac{ \pi} {4} +  \frac{ (k-1) \pi }{4k} ]  \nonumber  \\
&=&  [\frac{\pi}{4k},  \pi - \frac{\pi}{4k} ] \cup [- \pi +\frac{\pi}{4k}, -\frac{\pi}{4k}],
\eea
hence in that domain the spectrum of $1-g_kM_k$ lies out of the disk of center 0 and radius $\sin \frac{\pi}{4k}$.
Choosing $\epsilon =   \frac{\sin \frac{\pi}{4k} }{4\sqrt k}$, and assuming $\rho \le 1$, 
we have by \eqref{HS} $\Vert g_k M_k(Y) \Vert  \le \frac{1}{2}\sin \frac{\pi}{4k}$,
hence the spectrum of $ (1-g_kM_k (X) - i g_k M_k (Y) )$ lies out of the disk of center 0 and radius $\frac{1}{2}\sin \frac{\pi}{4k}$, and
\begin{equation}
\Vert (1-g_kM_k)^{-1}\Vert \le 2 [ \sin \frac{\pi}{4k} ]^{-1}. \label{resobound1}
\end{equation}
\qed

Since  $(1-g_kM_k)^{-1}$ has only two non trivial eigenvalues not equal to 1, this bound implies the same bound on the inverse square root
of its determinant $\det (1-g_kM_k)$, hence
\be \vert \exp\bigl[-\frac{1}{2}\Tr\ln(1-g_k M(\Psi))\bigr] \vert = \vert {\rm det}^{-1/2} (1-g_kM_k) \vert  \le 2 [ \sin \frac{\pi}{4k} ]^{-1} .
\ee
which means a uniform upper bound on the integrand in \eqref{interrepreal}.
Analyticity of $Z_k$ then follows by the standard argument based on Morera's theorem that a uniformly
convergent integral of an analytic integrand is analytic.

In fact there is clearly some margin still in the 
proof of Lemma \ref{analytlemma} and
with a litlle additional work we could check analyticity in a $D^k_\rho$ domain as done in the proof of Theorem \ref{mainbor}.

%

Finally the uniform estimates on the Taylor remainder at order $n$ can be obtained simply
by Taylor expanding $Z_k$ by the Taylor formula with integral remainder. This is similar to the proof of Theorem \ref{mainbor}
in Subsection \ref{borelsub} and left to the reader.

To complete the proof of Theorem \ref{maintheo}, one needs also tocheck that the perturbative expansion in $\lambda$ of this intermediate field representation is identical to the ordinary one, which is easy and left as exercise to the reader. Then by unicity of the Borel sum,
one concludes that the two integral representations \eqref{stand1} and \eqref{intfieldrep} must be equal. \qed


The LVE \cite{R1} as in \cite{Rivasseau:2010ke} 
is a technique to compute explicitly the logarithm of such partition functions and check its Borel sumability.
However we have not found yet how to adapt it to such intermediate field representations with
Gaussian imaginary integrals. 
The problem comes from the many replicas introduced 
by the LVE (one per vertex). Each of them should have its own small contour deformation 
and these deformations add up in a way which we do not know how to control as
the number $n$ of loop vertices tends to infinity.
Hence as a way out of this dilemma we give now, for the complex $k=3$ case, hence for the
$(\bar \phi \phi )^3$ model, another 
intermediate field representation, this time with \emph{bona fide} real Gaussian integrals rather than imaginary ones.

\section{Improved IF representation}
\label{improved}
\medskip

We return to \eqref{compkmodel} in the $k=3$ case, hence consider
\begin{equation}
Z^c_3(\lambda)= \int_{-\infty}^{+\infty}\int_{-\infty}^{+\infty} d \mu (\phi , \bar \phi ) e^{-\lambda(\bar\phi \phi)^{3}} .
\end{equation}
We split the interaction in two 
using a complex intermediate field $\sigma$ with normalized Gaussian measure $d\mu (\sigma)$
of covariance 1. The result is:
\begin{equation}
e^{- \lambda(\bar\phi \phi)^{3}}=\int{d\mu (\sigma)}e^{i\lambda^{1/2}(\bar\phi \phi)[ \bar \phi \sigma + \phi \bar \sigma]}.
\end{equation}
We  introduce complex conjugate fields intermediate fields $a$ and $\bar a$
 so that 
\begin{eqnarray}
e^{i\sqrt{\lambda}(\bar\phi \phi)[ \bar \phi \sigma + \phi \bar \sigma]} &=&\int d\mu(a) 
e^{ \sqrt i \lambda^{1/4} [\bar \phi\phi a  + (\bar \phi \sigma + \phi \bar \sigma) \bar a ]}.
\end{eqnarray}
The Gaussian integrals over $\phi$ and $\sigma$  can be explicitly performed, giving
\begin{eqnarray} \label{interrepreal}
Z^c_3(\lambda)&=&\int d\mu (\phi, \sigma, a) e^{ \sqrt i \lambda^{1/4} [\bar \phi\phi a  + (\bar \phi \sigma + \phi \bar \sigma) \bar a ]}\\
&=&\int d\mu(a) \det [1 -  \sqrt i \lambda^{1/4} \begin{pmatrix}
a &  \bar a \\ 
\bar a &  0
\end{pmatrix} ]^{-1}
\\
&=&\int d\mu(a)  f(a, \bar a ).
\label{pow}
\end{eqnarray}

Since $\det [1 -  \sqrt i \lambda^{1/4} \begin{pmatrix}
a &  \bar a \\ 
\bar a &  0
\end{pmatrix}] = 1 -  \sqrt i \lambda^{1/4} a - i \lambda^{1/2} \bar a^2 $, we have
\begin{equation} f(a, \bar a ) := (1 -  \sqrt i \lambda^{1/4} a - i \lambda^{1/2} \bar a^2 ) ^{-1}  = 
\sum_{n=0}^\infty (\sqrt i \lambda^{1/4} a + i \lambda^{1/2} \bar a^2)^n .
\end{equation}
Since any integral $\int d\mu(a) a^p \bar a ^q$ is zero unless $p=q$ we can in the functional integral \eqref{pow}
subsitute another \emph{perturbatively equivalent} function $ f_{\# a = \# \bar a}$ which simply discards, in the power series defining $f$,
any term not satisfying that constraint. This is a priori not justified non-perturbatively but will be justified \emph{a posteriori}
if we can obtain a Borel summable series by this process.
In our case $g = f_{\# a = \# \bar a}$ can be computed explicitly. More precisely
\begin{eqnarray}  f_{\# a = \# \bar a} &=& \sum_{n=3p} (-1)^p \lambda^{p}  (a \bar a)^{2p}  C^{3p}_{p}.
\end{eqnarray} 
The binomial coefficient
$C^{3p}_{p}= \frac{3p!}{p!(2p)!}$ is not far from the generalized Catalan number $C_{p}^{(3)} := \frac{1}{3p+1 }C^{3p+1}_{p} = 
\frac{1}{2p+1 }C^{3p}_{p}$.
We know that the \emph{alternating} generating function
\begin{equation}
h(x) = \sum_{p=0}^\infty (-1)^p C_{p}^{(3)} x^p \label{cardano}
\end{equation}
for such generalized Catalan numbers obeys the algebraic equation \cite{Bonzom:2011zz}
\begin{equation}
-xh^3(x) -h(x) +1 =0 .
\end{equation}
which  is soluble by radicals. 
More precisely defining  $ f_{\# a = \# \bar a} = g(\lambda^{1/2} a \bar a)$ and noting $u=\lambda^{1/2} a \bar a$, we have
\begin{eqnarray} g(u) = \sum_{p=0}^\infty (-1)^p  [u^{2}]^p C^{3p}_{p}  =  \sum_{p=0}^\infty ( 2p+1) (-1)^p (u^{2})^p C^{(3)}_{p}.
\end{eqnarray}
Hence the generating functions $g$ and $h$ are related through the simple equation
\begin{equation}
g (u) = \frac{d}{du}[ uh(u^2)] =  h(u^2) + 2 u^2 h'( u^2) .
\end{equation}

Returning to the solution of \eqref{cardano}, we define
\begin{equation}\Delta_\pm (y) :=  \biggl(\sqrt{1  +y} \pm \sqrt {y}   \biggr)^{1/3} = 1  \pm  \frac{1}{3}  \sqrt {y} + \frac{7y}{18}  \mp \frac{4y^{3/2}}{81}  + O(y^2)
\end{equation}
where we used that $(1+v)^{1/3} = 1+ v/3 - v^2/9 + 5 v^3/81 + O(v^4)$. The derivatives are easily computed as
\begin{equation}\Delta'_\pm (y) =  \frac{1}{6} \biggl((1  + y)^{-1/2} \pm  y^{-1/2} \biggr) \biggl(\sqrt{1  + y} \pm \sqrt {y}   \biggr)^{-2/3}
\end{equation}
Cardano's solution gives
\begin{eqnarray}
h(x) &=&\frac{\Delta_+( \frac{27}{4}  x)   - \Delta_{-}( \frac{27}{4}  x)}{\sqrt{3x}} = 1  -x   +  3x^2  + O(x^3)\nonumber \\
h'(x) &=&(3x)^{-3/2} [ \frac{81}{4} x(\Delta'_+ - \Delta'_{-}) -  \frac{3}{2}(\Delta_+ - \Delta_{-})] = -1 + 6x + O(x^2)\nonumber \\
\end{eqnarray}
where each $\Delta$ or $\Delta'$ is taken at $ \frac{27}{4}  x$. A nice simplification occurs in
\begin{equation}g (u) :=  h(u^2) + 2 u^2 h'( u^2) =  \frac{3^{5/2}}{2} u (\Delta'_+ - \Delta'_{-})\vert_{y= \frac{27}{4} u^2 },
\end{equation}
since the terms in $(\Delta_+ - \Delta_{-})$ disappear.
Let us compute  
\begin{equation} (\Delta'_\pm)\vert_{y= \frac{27}{4} u^2 } = \frac{1}{6u} \biggl(u (1  + \frac{27}{4} u^2 )^{-1/2} \pm  \sqrt { \frac{4}{27}}  \biggr) \biggl(\sqrt{1  +  \frac{27}{4} u^2 } \pm \sqrt { \frac{27}{4}  }u   \biggr)^{-2/3}.
\end{equation}
It leads to an explicit expression for $g$, analytic in a disk around $u=0$, namely
\begin{eqnarray}g(u) &=& \frac{1}{2}
\biggl[ \biggl(1+  \sqrt { \frac{27}{4}  }u(1  + \frac{27}{4} u^2 )^{-1/2}  \biggr) \biggl(\sqrt{1  +  \frac{27}{4} u^2 } + \sqrt { \frac{27}{4}  }u   \biggr)^{-2/3} 
\nonumber\\
&+& 
\biggl(1 - \sqrt { \frac{27}{4}  }u(1  + \frac{27}{4} u^2 )^{-1/2}  \biggr) \biggl(\sqrt{1  +  \frac{27}{4} u^2 } - \sqrt { \frac{27}{4}  }u   \biggr)^{-2/3}
\biggr]
\nonumber\\
&=& 1 - 3u^2 + 15 u^4 +\cdots 
\end{eqnarray}
Substituting in \eqref{pow} we end up with the \emph{improved}  IF representation with true (i.e. non imaginary) Gaussian measures

\medskip\noindent
{\bf Improved IF Representation} 
\begin{eqnarray} \label{interrepreal1}
Z(\lambda)&=&  \int d\mu (a) e^{S(a, \bar a)}, \ \quad S(a, \bar a)= S_1 + S_2   -\log 2  \\
%
S_1&=&  \log
\bigl[ \bigl(\sqrt{1  +  v^2 } + v  \bigr)^{1/3} + \bigl(\sqrt{1  + v^2 } - v   \bigr)^{1/3}\bigr]_{v=\sqrt { \frac{27 \lambda}{4}  }  a \bar a }\\
S_2&=&- \frac{1}{2} \log (1+ \frac{27 \lambda}{4}   (a \bar a)^2 )
\end{eqnarray}

This formula seems now well adapated to a loop vertex expansion because derivatives of $S$ remain bounded:

\noindent{\bf Lemma} {\it
The derivatives of $S$ with respect to any (strictly positive) number of $a$ and $\bar a$ variables are all uniformly bounded as $a \in {\mathbb C}$
and $\lambda$ runs in a suitable Borel-Leroy domain of order 2.}
\begin{equation} \vert \frac{\partial^p}{\partial a^p}   \frac{\partial^q}{\partial  \bar a^q}  S(a, \bar a) \vert
\le (p+q)! K^{p+q}  \vert \lambda\vert^{\frac{p+q}{4}} 
\end{equation}

\prf (Sketch)
The lemma is obviously proved if we prove it separately for $S_1$ and $S_2$.

For $S_2$ it is quite trivial as the Faa di Bruno forumla gives a sum of terms of the form 
$(1+ \frac{27 \lambda}{4}   (a \bar a)^2 )^{-r}  [ \frac{27 \lambda}{4} ]^s  a^t (\bar a)^u $
with $1 \le r \le p+q$, $s \ge (p+q)/4 $ and $t+u \le 3r$.

For $S_2$ it is slightly more complicated but a derivative acting on $\log D$ gives the factor $D'D^{-1}$
with $D:= \bigl(\sqrt{1  +  v^2 } + v  \bigr)^{1/3} + \bigl(\sqrt{1  + v^2 } - v   \bigr)^{1/3}$.
Deriving $\bigl(\sqrt{1  + v^2 } - v   \bigr)^{1/3}$ seems to create possibly an $a$ or $\bar a$ times
$D^{-1} [v(1  + v^2 )^{-1/2}-1  ] \bigl(\sqrt{1  + v^2 } - v   \bigr)^{-2/3} $, which would naively look unbounded.
But in fact at large $v$ positive $v(1  + v^2 )^{-1/2}$ tends to 1 and  there is a compensation 
$ [v(1  + v^2 )^{-1/2}-1  ] \bigl(\sqrt{1  + v^2 } - v   \bigr)^{-2/3}= v^{-2/3}[(1  + v^{-2} )^{-1/2}-1] [(1  + v^{-2} )^{1/2}-1]^{-2/3} \simeq  c v^{-4}  $.

\qed

The generalization of this improved representation to higher values of $k$ and the precise definition and convergence of the corresponding  loop vertex expansion
is however postponed to a future study.








%



\begin{thebibliography}{99} 

\bibitem{Rivasseau:2010ke} 
  V.~Rivasseau and Z.~Wang,
  ``Loop Vertex Expansion for Phi**2K Theory in Zero Dimension,''
  J.\ Math.\ Phys.\  {\bf 51}, 092304 (2010)
  doi:10.1063/1.3460320
  [arXiv:1003.1037 [math-ph]].



\bibitem{R1}
  V.~Rivasseau,
  ``Constructive Matrix Theory,''
  JHEP {\bf 0709} (2007) 008
  [arXiv:0706.1224 [hep-th]].

\bibitem{MR1}
  J.~Magnen and V.~Rivasseau,
  ``Constructive $\phi^4$ field theory without tears,''
  Annales Henri Poincare {\bf 9} (2008) 403
  [arXiv:0706.2457 [math-ph]].

\bibitem{Rivasseau:2009pi} 
  V.~Rivasseau,
  ``Constructive Field Theory in Zero Dimension,''
  Adv.\ Math.\ Phys.\  {\bf 2010}, 180159 (2010)
  doi:10.1155/2009/180159
  [arXiv:0906.3524 [math-ph]].

\bibitem{Rivasseau:2013ova} 
  V.~Rivasseau and Z.~Wang,
  ``How to Resum Feynman Graphs,''
  Annales Henri Poincare {\bf 15}, no. 11, 2069 (2014)
  doi:10.1007/s00023-013-0299-8
  [arXiv:1304.5913 [math-ph]].



\bibitem{MNRS}
  J.~Magnen, K.~Noui, V.~Rivasseau and M.~Smerlak,
  ``Scaling behaviour of three-dimensional group field theory,''
  Class.\ Quant.\ Grav.\  {\bf 26} (2009) 185012
  [arXiv:0906.5477 [hep-th]].


\bibitem{Gurau:2013pca} 
  R.~Gurau,
``The 1/N Expansion of Tensor Models Beyond Perturbation Theory,''
  Commun.\ Math.\ Phys.\  {\bf 330}, 973 (2014)
  doi:10.1007/s00220-014-1907-2
  [arXiv:1304.2666 [math-ph]].


\bibitem{Gurau:2014lua} 
  R.~Gurau and T.~Krajewski,
 ``Analyticity results for the cumulants in a random matrix model,''
  arXiv:1409.1705 [math-ph].

\bibitem{Rivasseau:2014bya} 
  V.~Rivasseau and Z.~Wang,
``Corrected loop vertex expansion for $\Phi_2^4$ theory,''
  J.\ Math.\ Phys.\  {\bf 56}, no. 6, 062301 (2015)
  doi:10.1063/1.4922116
  [arXiv:1406.7428 [math-ph]].


\bibitem{Delepouve:2014bma} 
  T.~Delepouve, R.~Gurau and V.~Rivasseau,
  ``Universality and Borel Summability of Arbitrary Quartic Tensor Models,''
  arXiv:1403.0170 [hep-th].


\bibitem{Delepouve:2014hfa} 
  T.~Delepouve and V.~Rivasseau,
  ``Constructive Tensor Field Theory: The $T^4_3$ Model,''
  arXiv:1412.5091 [math-ph].

%



\bibitem{Lahoche:2015yya} 
  V.~Lahoche,
  ``Constructive Tensorial Group Field Theory I:The $U(1)-T^4_3$ Model,''
  arXiv:1510.05050 [hep-th].

\bibitem{Lahoche:2015zya} 
  V.~Lahoche,
  ``Constructive Tensorial Group Field Theory II: The $U(1)-T^4_4$ Model,''
  arXiv:1510.05051 [hep-th].

\bibitem{Gurau:2013oqa} 
  R.~Gurau and V.~Rivasseau,
  ``The Multiscale Loop Vertex Expansion,''
  Annales Henri Poincare {\bf 16}, no. 8, 1869 (2015)
  doi:10.1007/s00023-014-0370-0
  [arXiv:1312.7226 [math-ph]].

\bibitem{Gurau:2011xp} 
  R.~Gurau and J.~P.~Ryan,
  ``Colored Tensor Models - a review,''
  SIGMA {\bf 8}, 020 (2012)
  doi:10.3842/SIGMA.2012.020
  [arXiv:1109.4812 [hep-th]].



\bibitem{Sok} 
 A.~D.~Sokal,
  ``An Improvement Of Watson's Theorem On Borel Summability,''
  J.\ Math.\ Phys.\  {\bf 21}, 261 (1980).



\bibitem{BK} D. Brydges and T. Kennedy, 
Mayer expansions and the Hamilton-Jacobi equation,
Journal of Statistical Physics, {\bf 48}, 19 (1987).



\bibitem{AR1}
  A.~Abdesselam and V.~Rivasseau,
  ``Trees, forests and jungles: A botanical garden for cluster expansions,''
  arXiv:hep-th/9409094.

\bibitem{Bonzom:2011zz} 
  V.~Bonzom, R.~Gurau, A.~Riello and V.~Rivasseau,
  ``Critical behavior of colored tensor models in the large N limit,''
  Nucl.\ Phys.\ B {\bf 853}, 174 (2011)
  doi:10.1016/j.nuclphysb.2011.07.022
  [arXiv:1105.3122 [hep-th]].

\end{thebibliography}
\end{document}